\documentclass{appolb}
\usepackage{graphicx}
\usepackage{cite}
\usepackage{subfig}
\usepackage{hyperref}
\usepackage{amssymb}
\begin{document}
\eqsec  
\title{OFF-SHELL $t\bar{t}b\bar{b}$ PRODUCTION AT THE LHC: \\ QCD CORRECTIONS, THEORY UNCERTAINTIES \\ AND $b$-JET DEFINITIONS
\thanks{Presented at the XLIV International Conference of Theoretical Physics "Matter To The Deepest" (virtual edition), September 15-17, 2021}
} 
\author{\vspace{0.5cm} Giuseppe Bevilacqua
\address{ELKH-DE Particle Physics Research Group, University of Debrecen, H-4010 Debrecen, PBox 105, Hungary}
}

\maketitle
\begin{abstract}
We present state-of-the-art predictions for off-shell $t\bar{t}b\bar{b}$ production with di-lepton decays at the LHC with $\sqrt{s}=13$ TeV. Results are accurate at NLO in QCD and include all resonant and non-resonant diagrams, interferences and finite-width effects for top quarks and $W$ bosons. We discuss the impact of QCD corrections and assess theoretical uncertainties  from scale and PDF dependence at the integrated and differential level. Additionally we investigate the size of contributions induced by initial-state $b$ quarks to the NLO cross section.
\end{abstract}

%---------------------------
\section{Introduction}
%---------------------------

Since the initial announcement of a newly discovered scalar particle in 2012, extensive efforts have been made at the Large Hadron Collider to confirm its position as the Standard Model (SM) Higgs boson. One of the crucial tests is to probe whether the boson couples with the fundamental fermions via Yukawa interaction and proportionally to the fermion mass. As the heaviest particle in the SM, the top quark is the most sensitive probe. One of the most interesting processes that can be studied at LHC in this context is $pp \to t\bar{t}H \to t\bar{t}b\bar{b}$, which offers a direct probe of the Higgs coupling to top quark and benefits from the relatively large branching ratio of the $H\to b\bar{b}$ decay. The precise measurement of the $t\bar{t}H(H\to b\bar{b})$ signal, however, presents a number of challenges. In the first place one has to face with the so-called \textit{combinatorial background}, related to the ambiguity of identifying $b$-jets as decay products of the Higgs boson or top quarks. The combinatorial background leads to a sensible smearing of the Higgs boson peak in the $b\bar{b}$ invariant mass distribution. Another very important issue is related to the presence of large SM  backgrounds which impact the sensitivity of the $t\bar{t}H$ signal extraction. Among various sources of background, the QCD process $pp \to t\bar{t}b\bar{b}$ stands out in that it features the same final state composition as the signal. For this reason it plays the role of \textit{irreducible} background to $t\bar{t}H(H\to b\bar{b})$. Interestingly, the $t\bar{t}b\bar{b}$ process represents the main background for final states with at least four $b$-tagged jets (see \textit{e.g.} \cite{CMS:2018hnq}). Measurements of the $t\bar{t}b\bar{b}$ cross section at $\sqrt{s} = 13$ TeV have been reported by both the ATLAS and CMS collaborations \cite{CMS:2017xnm,ATLAS:2018fwl,CMS:2020grm}. A slight excess has been observed with respect to the results from NLO+PS simulations, which points to the need for improved theoretical modelling.

The process of $t\bar{t}b\bar{b}$ hadroproduction has been increasingly investigated at NLO accuracy since more than a decade now. The first NLO studies were carried out in the picture of stable top quarks \cite{Bredenstein:2009aj,Bevilacqua:2009zn,Bredenstein:2010rs}, followed by a number of analyses which interfaced the NLO calculation to parton showers \cite{Garzelli:2014aba,Cascioli:2013era,LHCHiggsCrossSectionWorkingGroup:2016ypw,Bevilacqua:2017cru,Jezo:2018yaf}. These results were complemented by other studies providing accessory information on $t\bar{t}b\bar{b}$ from different viewpoints, such as  investigations of the $t\bar{t}b\bar{b}$/$t\bar{t}jj$ cross section ratio \cite{Bevilacqua:2014qfa} and the analysis of $t\bar{t}b\bar{b}$ production in association with a light jet \cite{Buccioni:2019plc}. In all the studies mentioned above, top quarks were considered on-shell. Decays, when present, were described at LO accuracy including spin correlations. It is only very recently that the first \textit{off-shell} predictions, based on a complete NLO QCD calculation at fixed perturbative order, have started to  appear \cite{Denner:2020orv,Bevilacqua:2021cit}. In this proceedings we summarize the results of one of these studies, as presented in \cite{Bevilacqua:2021cit}.

%--------------------------------------------
\section{Details of the calculation}
\label{sec:setup}
%--------------------------------------------

The present study pertains $t\bar{t}b\bar{b}$ production with di-lepton top-quark decays at the LHC with center-of-mass energy $\sqrt{s} = 13$ TeV. To be more precise, we compute the process $pp \to e^+\nu_e \, \mu^- \bar{\nu}_\mu \, b\bar{b} \, b\bar{b} + X$ at NLO QCD accuracy. In the following we will also refer to the latter process as to $t\bar{t}b\bar{b}$ for ease of notation. It should be clear though, that all resonant and non-resonant Feynman diagrams, interferences and finite-width effects at the perturbative order $\mathcal{O}(\alpha^4 \alpha_s^5)$ are included in the calculation. 

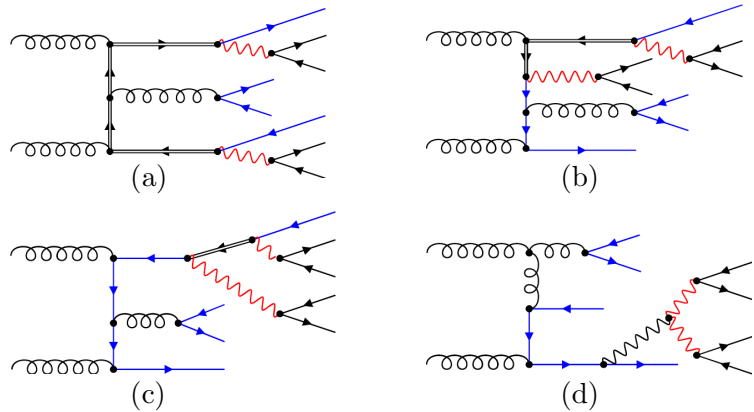
\begin{figure}[h!tb]
\centerline{
\put(-133,0){\includegraphics[width=0.33\textwidth]{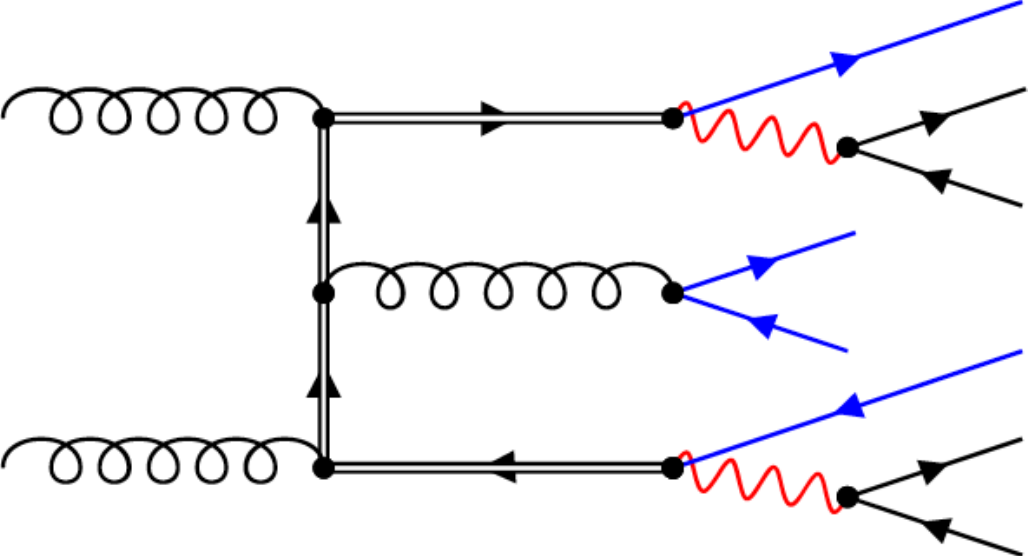}}
\put(-88,-3){(a)}
\hspace{0.04\textwidth}
\put(23,9){\includegraphics[width=0.33\textwidth]{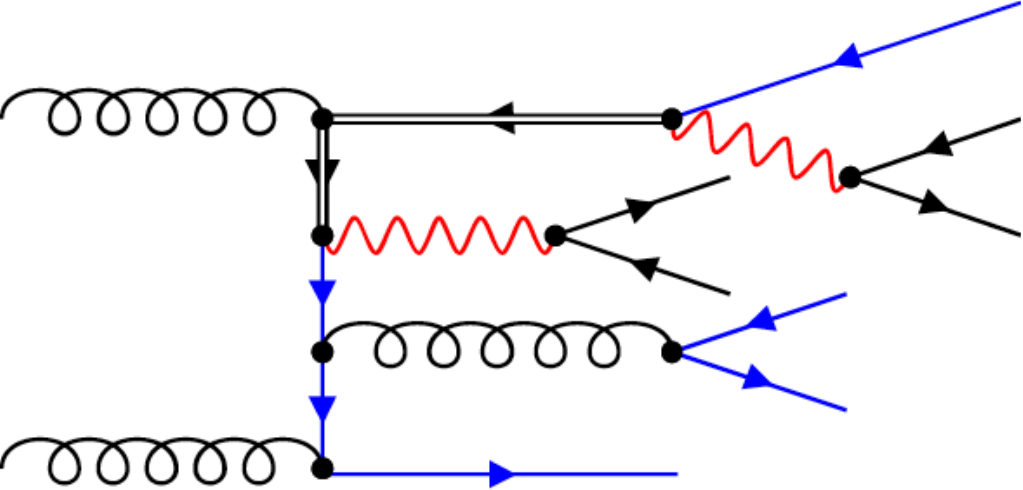}}
\put(73,-3){(b)}
}
\medskip 
\centerline{
\put(-133,0){\includegraphics[width=0.34\textwidth]{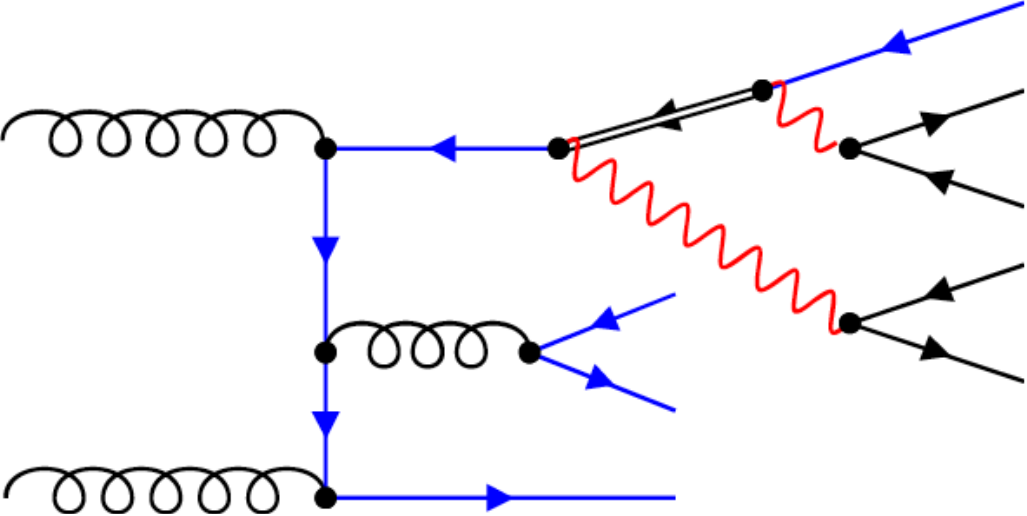}}
\put(-88,-11){(c)}
\hspace{0.04\textwidth}
\put(23,0){\includegraphics[width=0.34\textwidth]{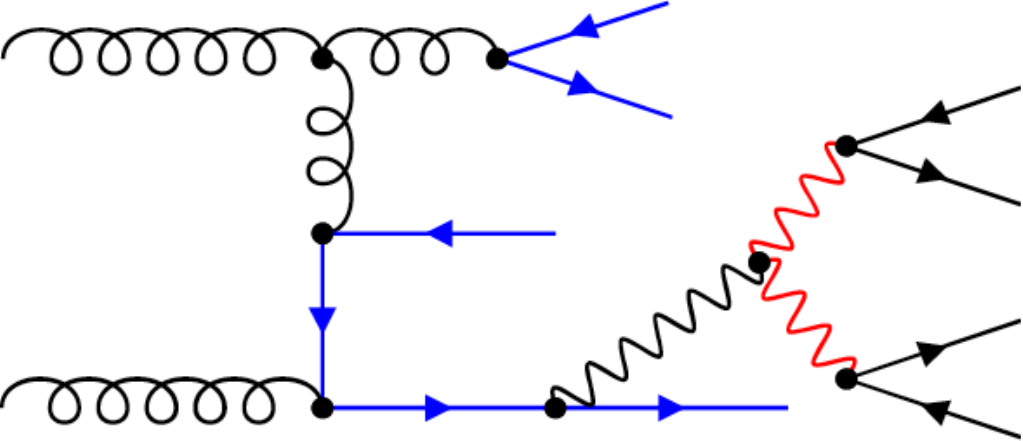}}
\put(73,-11){(d)}
}
\medskip
\caption{Representative Feynman diagrams contributing to the process $gg \to e^+ \nu_e \, \mu^- \bar{\nu}_ \mu \, b \bar{b} \, b \bar{b} $ at Born level: \textit{double-resonant} (a), \textit{single-resonant} (b,c), \textit{non-resonant} (d). The double-lined propagators represent top quarks, the red propagators $W$ bosons and the blue lines bottom quarks.}
\label{Fig:diagrams}
\end{figure}

It is instructive to inspect a few representative Feynman diagrams describing the dominant $gg$ channel, see Figure \ref{Fig:diagrams}. A quick look at the intermediate propagators is sufficient to get the rich structure of resonances induced by the dynamics of the process. On the one hand (case (a) in Fig.\ref{Fig:diagrams}) there are diagrams which enhance double top-quark resonances in the final state. Other diagrams (b,c) induce "$tW$"-like signatures which are characterised by a single top-quark resonance. Finally (d) there are diagrams which do not encompass top-quark propagators at all, yet they introduce additional multi-boson resonances. The relative importance of \textit{double-}, \textit{single-} and \textit{non-resonant} (from top-quark viewpoint) contributions depends on the actual setup of the analysis. When sufficiently inclusive kinematical cuts are adopted, double-resonant contributions describe the bulk of the fiducial cross section \cite{Fadin:1993kt}. Under this assumption the Narrow Width Approximation (NWA), based on the limit $\Gamma_t/m_t \to 0$, can be used to obtain results very close to the full calculation while reducing dramatically the computational burden (since only double-resonant diagrams are retained). On the other hand, the reliability of the NWA becomes questionable if more restrictive cuts are adopted (for example when $t\bar{t}b\bar{b}$ is a background process that one would like to  suppress). In the latter case the single- and non-resonant contributions \-- that we overall refer under the name "\textit{off-shell effects}" for brevity \-- might play a more prominent role and should be incorporated in the calculation in order to obtain more accurate predictions. In any case, it appears clearly that $t\bar{t}b\bar{b}$ is a genuine multi-scale process and that $m_t$ does not necessarily set the most natural scale. We will comment further on this point in Section \ref{sec:results}.

The present analysis requires final states with at least four $b$-jets, two charged leptons and missing $p_T$. Jets are defined according to the anti-$k_T$ algorithm \cite{Cacciari:2008gp} with resolution parameter $R=0.4$. The following kinematical cuts are imposed:
\begin{equation}
p_{T,\,\ell}>20 ~{\rm GeV}\,,  \;\;\;\;  p_{T,\,b}>25  ~{\rm GeV}\,,  \;\;\;\;  |y_\ell|<2.5\,,  \;\;\;\;  |y_b|<2.5   \,, 
\end{equation}
where $\ell = \mu^-,e^+$. Neither  the extra jet nor the missing transverse momentum have any restriction imposed. We consider this set of kinematical cuts to be rather inclusive. Following the PDF4LHC recommendations for applications at the LHC Run II \cite{Butterworth:2015oua}, we adopt the latest fits by several groups: NNPDF3.1 \cite{NNPDF:2017mvq}, MMHT2014 \cite{Harland-Lang:2014zoa} and CT18NLO \cite{Hou:2019efy}. It is worth to mention that $b$ quarks are treated massless in our calculation, \textit{i.e.} we work in the 5-Flavor Number scheme. Further details of our calculational setup can be found in Ref.\cite{Bevilacqua:2021cit}. 

On the technical side, all results have been obtained with the help of the Monte Carlo framework \textsc{Helac-Nlo} \cite{Bevilacqua:2011xh}, which consists of \textsc{Helac-1loop} \cite{Ossola:2007ax,vanHameren:2010cp,vanHameren:2009dr} and \textsc{Helac-dipoles} \cite{Czakon:2009ss,Bevilacqua:2013iha}. Phase space integrations are performed with \textsc{Kaleu}  \cite{vanHameren:2010gg}. Our results are available in the form of either Les Houches Event Files \cite{Alwall:2006yp} or \textsc{Root} Ntuples \cite{Antcheva:2009zz} which might be directly used in connection with experimental analyses. The events are equipped with matrix-element and PDF information to allow on-the-fly reweighting for different scales and PDF sets \cite{Bern:2013zja}. The Ntuples are processed by a in-house C++ analysis framework, \textsc{Heplot}, to obtain predictions for any infrared-safe observable, scale/PDF setup, using customized kinematical cuts.

We have performed a number of consistency checks for our calculation, both internally and in connection with published results. We successfully reproduced the results of Ref.\cite{Denner:2020orv} for various differential cross sections, finding very good agreement in all cases.

%-------------------------------------------------------------------
\section{Numerical results}
\label{sec:results}
%-------------------------------------------------------------------

We begin our discussion with the analysis of the fiducial cross section, \textit{i.e.} the total cross section as obtained via integration over the fiducial phase space. The purpose is to monitor to what extent different choices of renormalisation and factorisation scales impact this quite inclusive observable. To this end we compare predictions obtained using two functional forms for the scales: (i) $\mu_1 = m_t$ and (ii) $\mu_2 = H_T/3$, where
\begin{equation}
H_T = p_T(b_1) + p_T(b_2) + p_T(b_3) + p_T(b_4) + p_T(e^+) + p_T(\mu^-) + p_T^{miss} \,.
\end{equation}
Here $b_1,b_2,b_3,b_4$ denote the hard $b$-tagged jets in the final state, ordered in $p_T$.
$\mu_1$ is an example of \textit{fixed} scale while $\mu_2$ is \textit{dynamical}, namely its value varies from event to event. The first scale is expected to provide an adequate description of the process particularly in the vicinity of the $t\bar{t}$ production threshold. As anticipated in section \ref{sec:setup}, the genuine multi-scale nature of $t\bar{t}b\bar{b}$ is such that other mechanisms can potentially play an important role away from the threshold. Thus the second scale choice might perform better particularly in the high energy regime. In Table \ref{tab:total_xsec} we report our findings for the fiducial cross sections at LO and NLO, as obtained using the NNPDF3.1 PDF set. We observe pretty large QCD corrections for both scales, in agreement with the earliest findings on $t\bar{t}b\bar{b}$ at NLO \cite{Bredenstein:2009aj,Bevilacqua:2009zn,Bredenstein:2010rs}. Also, the central values of the NLO cross sections for the two scale choices look very similar. So do the estimated scale uncertainties, which are of the order of 20\%. The observed stability of the fiducial NLO cross section upon different scales is somewhat expected given the rather inclusive fiducial cuts that we are considering. However, where a dynamical scale choice such as $\mu = H_T/3$ shows its strength is at the differential level: in line with our findings from earlier studies of off-shell $t\bar{t}+X$ production ($X=j,\gamma,W^\pm,Z(\to \nu\bar{\nu})$) \cite{Bevilacqua:2016jfk,Bevilacqua:2018woc,Bevilacqua:2019cvp,Bevilacqua:2020pzy}, we have observed that various dimensionful observables (\textit{e.g.} transverse momentum and invariant mass distributions) benefit from the $H_T$-based scale choice particularly in the high energy tails. On a bin-by-bin basis, $K$-factors appear flatter and the NLO uncertainty bands fit more nicely into the LO ones. We consider the latter feature to be an indication of better perturbative convergence of our predictions, thus in the rest of the discussion we will concentrate on results obtained with the scale $\mu=H_T/3$. 
\begin{table}[h!tb]
\hspace{0.5cm}
\begin{tabular}{|c|c|c|c|}
  \hline
 $\rm{Scale}$  &  $\sigma_{\rm{LO}} \, \rm{[fb]}$  &  $\sigma_{\rm{NLO}} \, \rm{[fb]}$  &  $K = \sigma_{\rm{NLO}}/\sigma_{\rm{LO}}$  \\[0.1cm]
  \hline \hline 
  $\mu_0 = {m_t}$   &    $6.998^{+4.525 \,(65\%)}_{-2.569 \, (37\%)}$    &   $13.24^{+2.33 \,(18\%)}_{-2.89 \, (22\%)}$   &   $1.89$   \\[0.22cm]
  \hline 
  $\mu_0 = {H_T/3}$   &   $6.813^{+4.338 \,(64\%)}_{-2.481 \, (36\%)}$   &   $13.22^{+2.66 \,(20\%)}_{-2.95 \, (22\%)}$    &   $1.94$   \\[0.12cm]
  \hline
 \end{tabular} \smallskip
\caption{ \label{tab:total_xsec} 
LO and NLO fiducial cross sections for $pp \to e^+\nu_e\,\mu^-\bar{\nu}_\mu\,b\bar{b}\,b\bar{b}\, + X$ at the LHC with $\sqrt{s} = 13$ TeV, based on the {NNPDF3.1} PDF set. The errors denote scale uncertainties. In the last column the $K$-factor is shown.
}
\end{table}

\begin{figure}[h!tb]
\centerline{
\includegraphics[width=0.52\textwidth,height=0.43\textwidth]{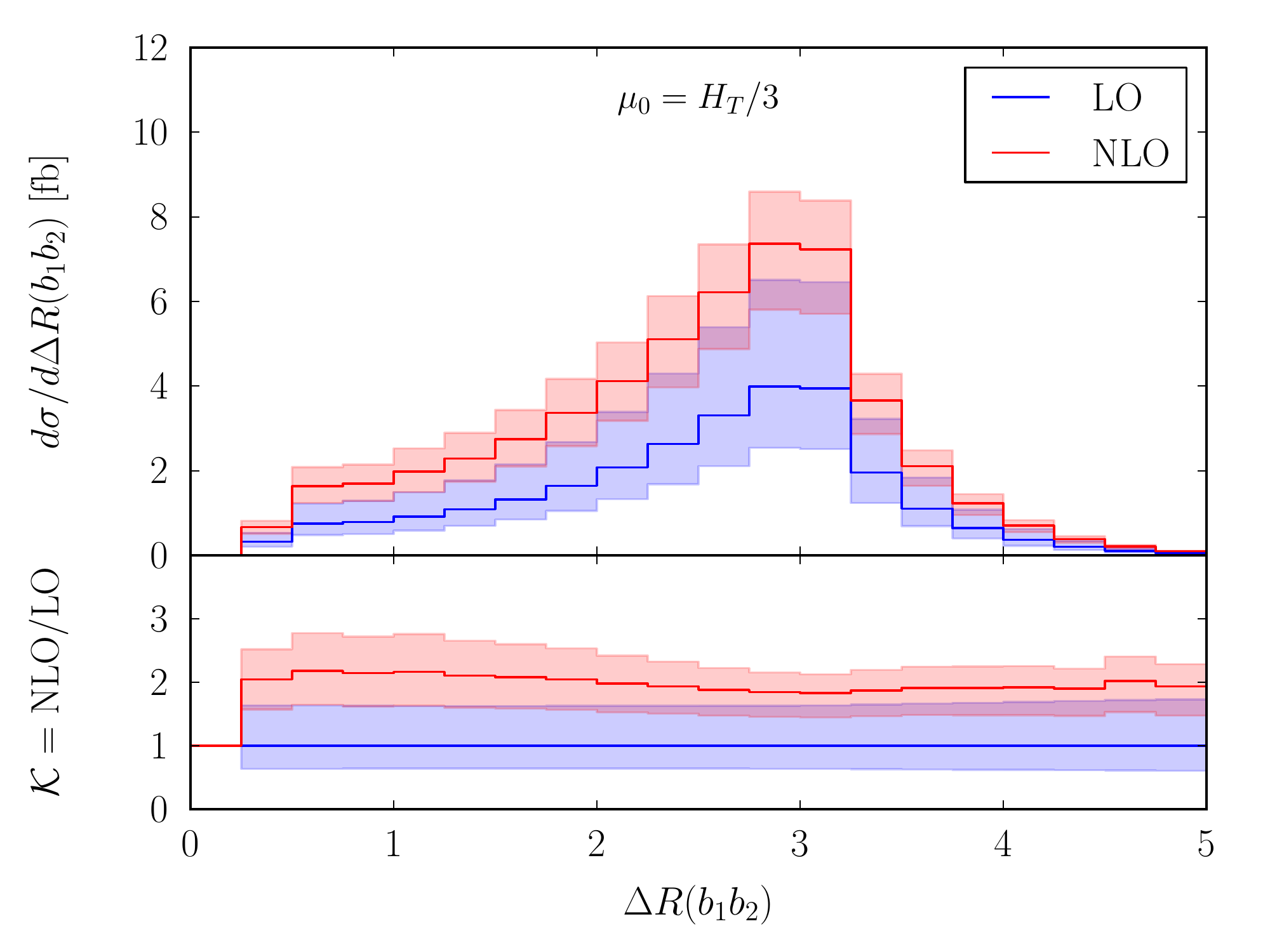}
\includegraphics[width=0.52\textwidth,height=0.43\textwidth]{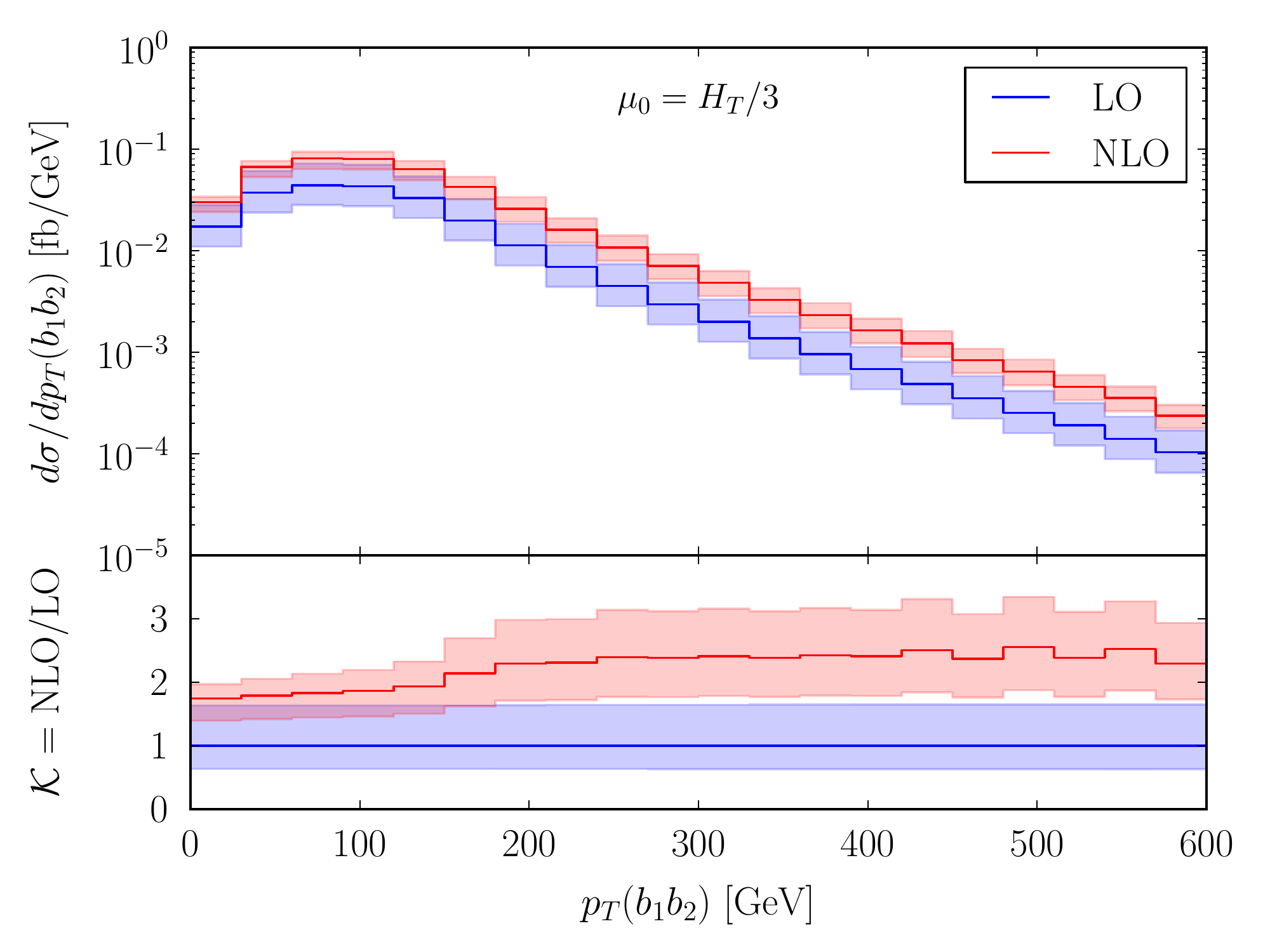}
}
\caption{LO and NLO differential cross sections as a function of $\Delta R(b_1b_2)$ and $p_T(b_1b_2)$ (defined in the text). Results are based on the scale choice $\mu=H_T/3$ and on the {NNPDF3.1} PDF set. The bands denote scale uncertainties.
}
\label{fig:qcd_corrections}
\end{figure}

Let us now analyse the impact of QCD corrections and scale uncertainties at the differential level. In Fig.\ref{fig:qcd_corrections}, two observables related to the kinematics of $b$-jets are shown: the $\Delta R$ separation and the transverse momentum of the $b_1b_2$ system. Looking at $\Delta R(b_1b_2)$, we note that the two $b$-jets are generated mostly in back-to-back configurations. In contrast, the remaining two hard $b$-jets present in the final state ($b_3,b_4$) are found to peak at smaller $\Delta R$ values\cite{Bevilacqua:2021cit}. The behaviour of the $b_3b_4$ pair is closer to the expectations from a $b$-jet pair originated by gluon splitting, thus it is tempting to consider $b_3,b_4$ to be the "prompt" $b$-jets in $t\bar{t}b\bar{b}$ production and ascribe $b_1,b_2$ to the decays of top quarks. Furthermore, looking at $p_T(b_1b_2)$, we note that the $K$-factor is far from being constant in the observed range and reaches up to $\mathcal{O}(2.5)$ for $p_T(b_1b_2) \gtrsim 150$ GeV. The size of the LO and NLO scale bands becomes comparable in the tail of the distribution. This is a genuine effect from dominant real-radiation contributions.

\begin{figure}[h!tb]
\centerline{
\includegraphics[width=0.5\textwidth,height=6.8cm]{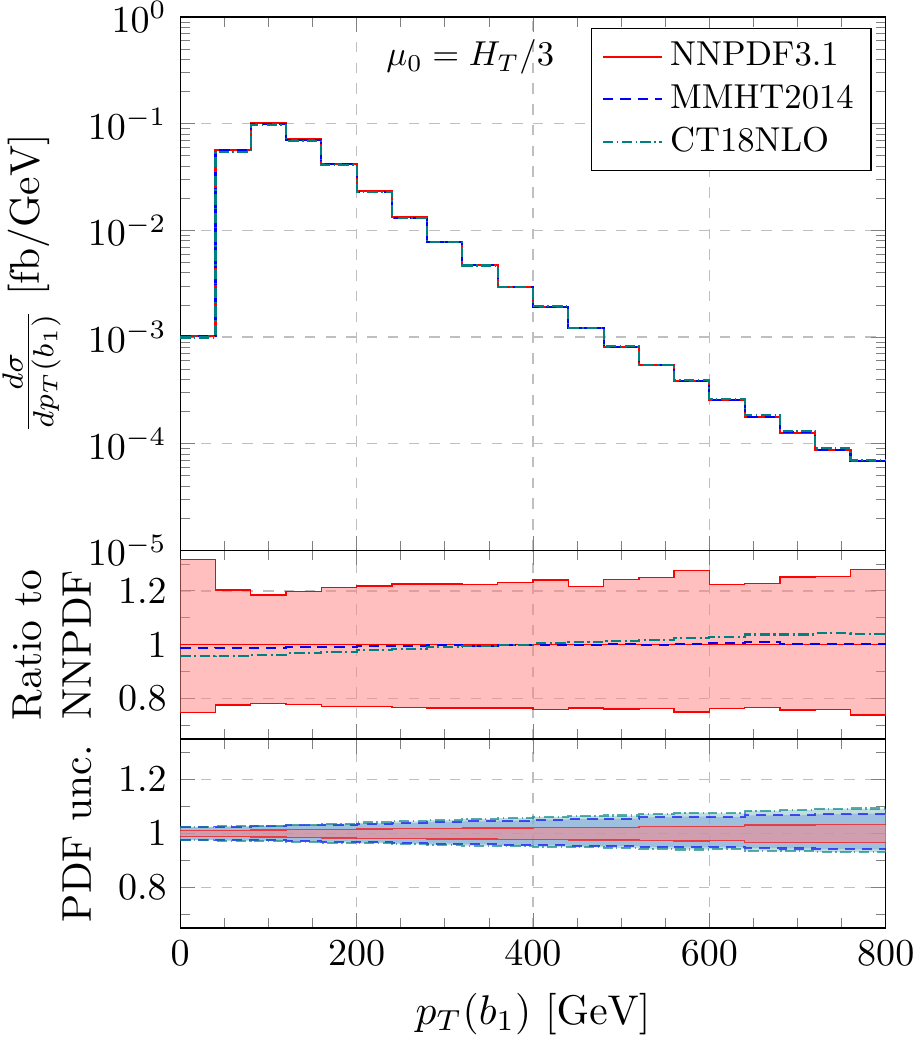}
\includegraphics[width=0.5\textwidth,height=6.8cm]{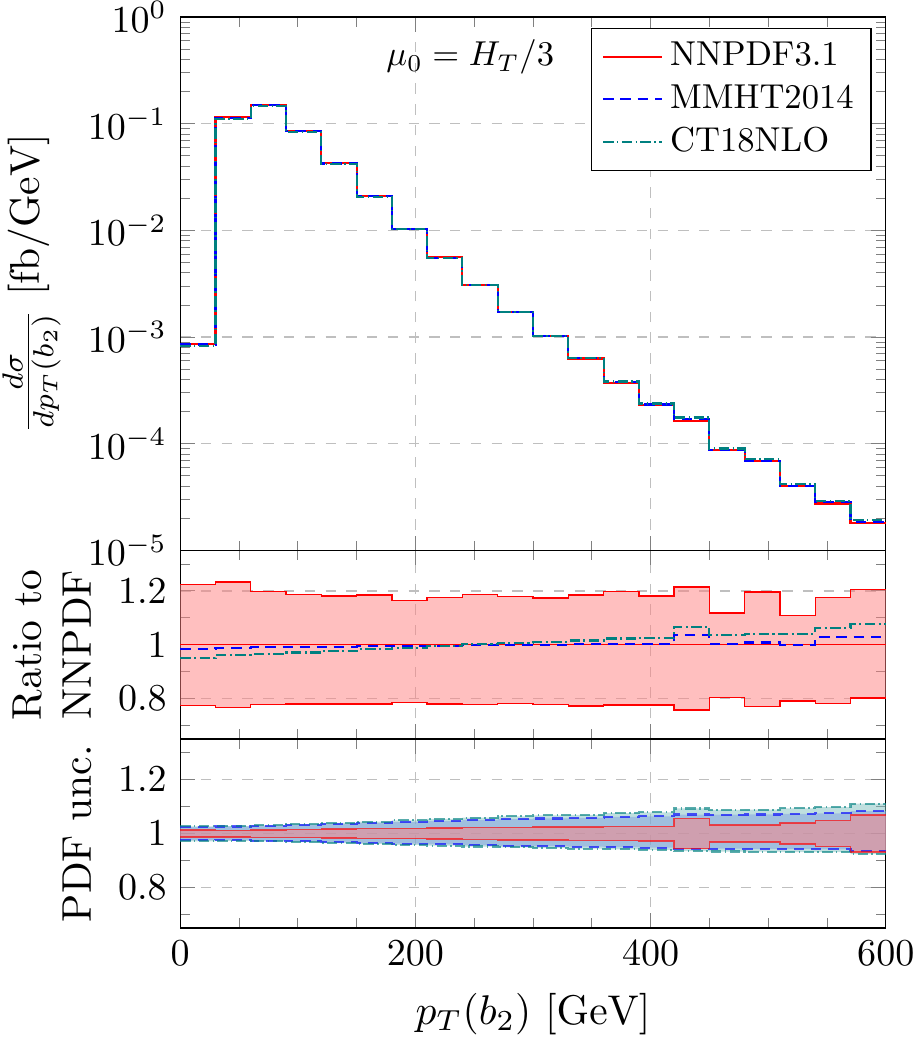}
}
\caption{NLO differential cross section as a function of $p_T(b_1)$ and $p_T(b_2)$ (defined in the text), for three different input PDF sets. The bands in the middle(lower) inset denote scale(PDF) uncertainties.}
\label{fig:PDF_uncertainties}
\end{figure}

We complete the analysis of the NLO theoretical uncertainties by assessing the PDF dependence at the differential level, see Fig. \ref{fig:PDF_uncertainties}. The three bands shown in the lower inset correspond to the internal uncertainties of the various PDF sets. The latter should be compared with the band reported in the middle inset, which is the estimated scale uncertainty for the reference setup $\mu=H_T/3$. We note that PDF uncertainties amount to few percents in the bulk of the distributions and can reach up to $\mathcal{O}(10\%)$ in tails, yet are well below the scale uncertainties. We have extensively checked that similar conclusions hold for other dimensionful and dimensionless observables examined in our study\cite{Bevilacqua:2021cit}.

\begin{figure}[h!tb]
\centerline{
\includegraphics[width=0.5\textwidth]{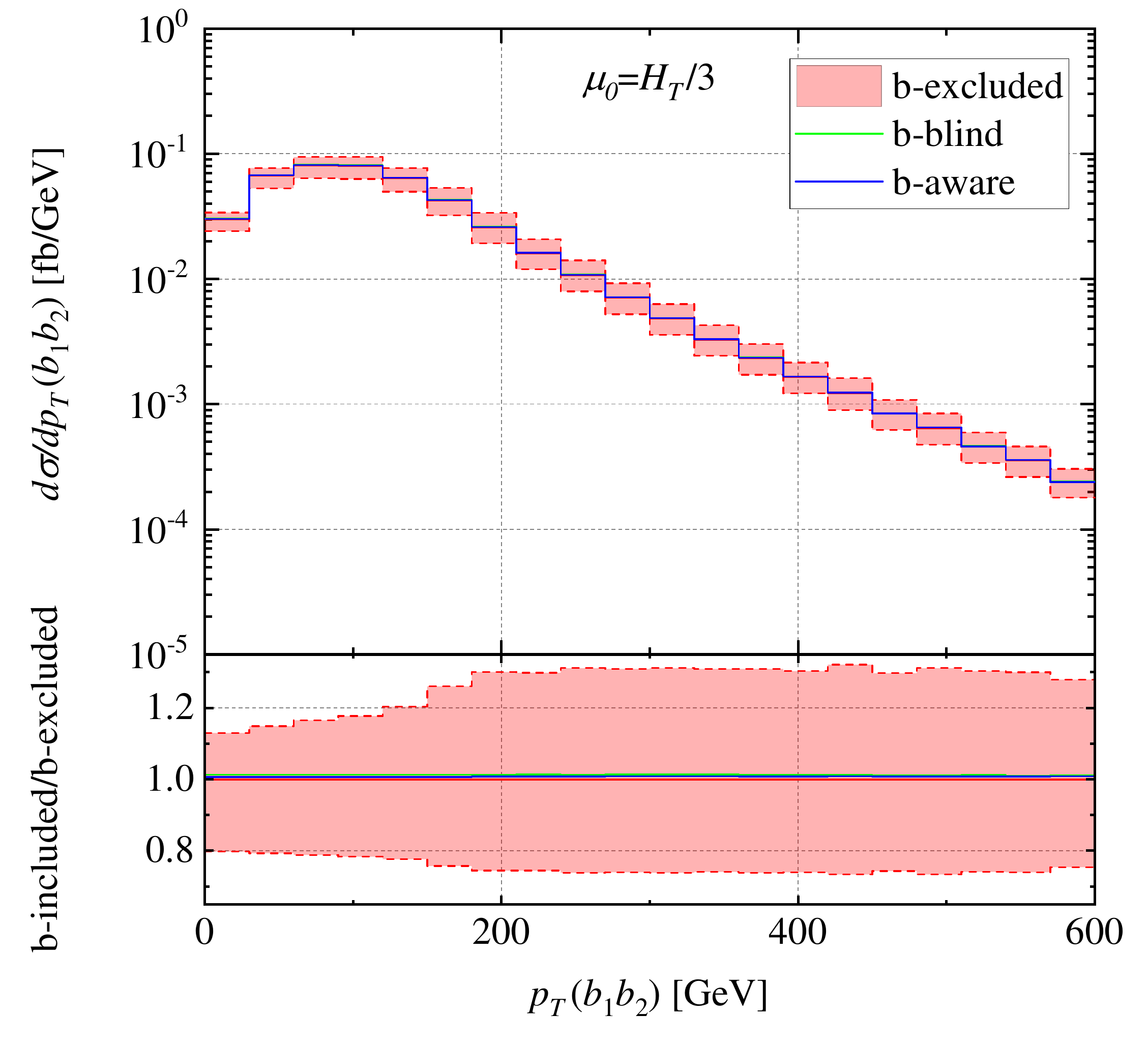}
\includegraphics[width=0.5\textwidth]{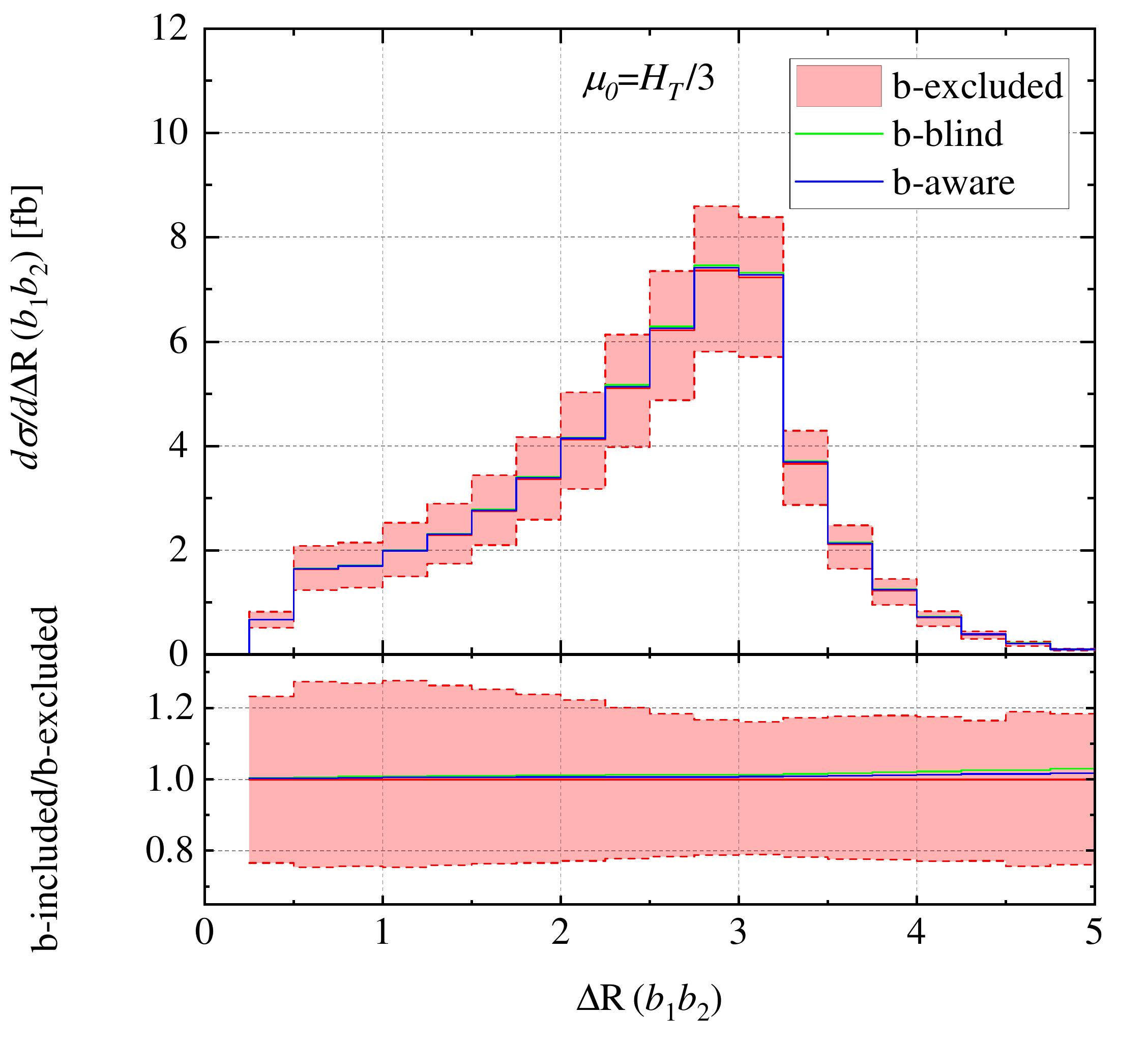}
}
\caption{NLO differential cross sections as a function of $p_T(b_1b_2)$ and $\Delta R(b_1b_2)$ (defined in the text). Full NLO predictions without and with initial-state $b$ contributions (for charge-aware and charge-blind $b$ tagging) are compared. The bands denote scale uncertainties.}
\label{fig:includeb_vs_excludeb}
\end{figure}

\begin{figure}[h!tb]
{
\hspace{2.2cm}
\includegraphics[width=0.65\textwidth]{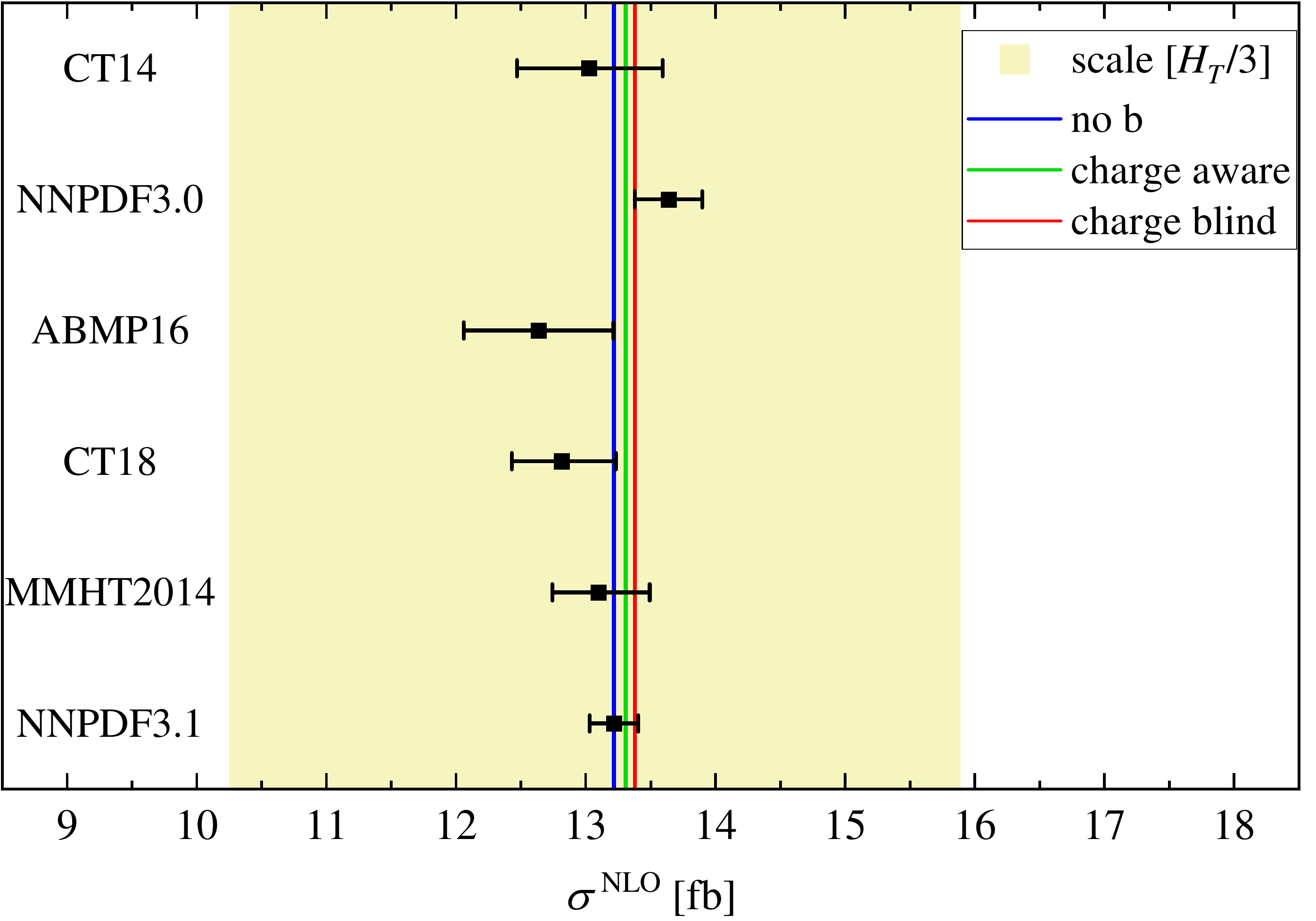}
}
\caption{NLO fiducial cross sections for $pp \to e^+\nu_e \mu^- \, \bar{\nu}_\mu \, b\bar{b} \, b\bar{b} + X$ at the LHC with $\sqrt{s} = 13$ TeV. The yellow band shows the scale uncertainty for $\mu_0 = H_T/3$. The error bars represent the internal PDF uncertainties. The vertical lines denote central values of the NLO cross section for the NNPDF3.1 set.
}
\label{fig:scale_pdf_uncertainties_paleyellow}
\end{figure}

In the next step we address the question of how important are contributions induced by initial-state $b$-quarks to the NLO cross section. The latter contributions comprise $gb$, $g\bar{b}$, $b\bar{b}$, $bb$ and $\bar{b}\bar{b}$ channels. The expectation is that they are globally suppressed by small $b(\bar{b})$-parton luminosities. Neglecting these channels can help to simplify the bookkeping of the NLO calculation, on condition that they do not originate unexpectedly relevant effects in some region of the phase space. We check that this is indeed the case considering two different approaches of $b$-tagging, that we name "\textit{charge-aware}" and "\textit{charge-blind}". The basic difference is that the charge-aware tagging is sensitive to the flavour \textit{and} to the charge of jets, whereas in the charge-blind case only the flavour information is available. The recombination rules are slightly different in the two cases:
\begin{eqnarray}
\mbox{charge-aware: } \;\; bg \to b, \;\;\; \bar{b}g \to \bar{b}, \;\;\;  b\bar{b} \to g, \;\;\;  bb \to b,  \;\;\;  \bar{b}\bar{b} \to \bar{b}  \\
\mbox{charge-blind: } \;\; bg \to b, \;\;\; \bar{b}g \to \bar{b}, \;\;\;  b\bar{b} \to g, \;\;\;  bb \to g,  \;\;\;  \bar{b}\bar{b} \to g 
\end{eqnarray}
In practice only the $bb$ and $\bar{b}\bar{b}$ rules differ. Hence the two schemes represent equally infrared-safe variants that can be used for our NLO calculation. Let us further note that, in the charge-aware scheme, contributions from $bb$ and $\bar{b}\bar{b}$ initial states are absent because we require final states with at least two $b$- and two $\bar{b}$-jets. Our findings for the fiducial cross section at LO and NLO, based on $\mu=H_T/3$ and on the NNPDF3.1 set, are the following:
\begin{eqnarray}
\sigma_{\rm{no b}}^{\rm{LO}} = 6.813(3) \, \rm{fb}, ~~~~ \sigma_{\rm{aware}}^{\rm{LO}} = 6.822(3) \, \rm{fb}, ~~~~  \sigma_{\rm{blind}}^{\rm{LO}} = 6.828(3) \, \rm{fb}, \nonumber \\[0.15cm]
\sigma_{\rm{no b}}^{\rm{NLO}} = 13.22(3) \, \rm{fb}, ~~~~ \sigma_{\rm{aware}}^{\rm{NLO}} = 13.31(3) \,\rm{fb}, ~~~~  \sigma_{\rm{blind}}^{\rm{NLO}} = 13.38(3) \, \rm{fb}.  \nonumber
\end{eqnarray}
The subscript "$\rm{nob}$" indicates results where contributions from $b$-initiated process have been neglected, whereas "$\rm{blind}$" and "$\rm{aware}$" denote the full results obtained using the two $b$-tagging approaches described above. One can see that the effects of initial-state $b$ contributions amount to $0.2\%$ at LO, and reach up to $1\%$ at NLO (mainly due to the opening of $gb$ channels). As shown in Fig. \ref{fig:includeb_vs_excludeb}, similar conclusions hold at the  differential level. To put things in perspective, it is useful to compare the shift induced by initial-state $b$ contributions to the size of other theoretical uncertainties: this is shown in Fig.\ref{fig:scale_pdf_uncertainties_paleyellow}. To have a broader view we have included here results based on additional PDF sets: CT14 \cite{Dulat:2015mca}, NNPDF3.0 \cite{Ball:2014uwa} and ABMP16 \cite{Alekhin:2018pai}. The conclusion is that initial-state $b$ quark contributions can be safely neglected in our process.

To conclude the discussion, we have been able to compare our predictions for the fiducial cross section with the recent measurement by the ATLAS collaboration in the $e\mu$ top-quark decay channel \cite{ATLAS:2018fwl}, using the same cuts of the ATLAS analysis:
\begin{equation}
\sigma^{\scriptsize \textsc{ATLAS}}_{e\mu+4b} = (25 \pm 6.5) ~ \rm{fb} \,,
\qquad \quad
\sigma^{\scriptsize \textsc{Helac-Nlo}}_{e\mu+4b} = (20.0 \pm 4.3) ~ \rm{fb} \,.
\end{equation}
The prediction is in good agreement with the experimental measurement at the present accuracy.

%---------------------------
\section{Conclusions}
%---------------------------

We have presented state-of-the-art predictions for $pp \to t\bar{t}b\bar{b}$ production with di-lepton decays at $\sqrt{s}=13$ TeV, including complete finite-width and non-resonant effects for top-quark and $W$ decays at NLO QCD accuracy. Large and positive QCD corrections of the order of $90\%$ have been observed when looking at the integrated fiducial cross section. The latter  reach even larger values differentially. We extensively examined the NLO theoretical uncertainties related to scale and PDF variation, being able to quantify them at the level 20\% and $1\%-3\%$ respectively. Thus, scale dependence is the dominant source of NLO uncertainty. Finally we quantified the impact of $b$-initiated subprocesses on the NLO cross section, showing that they can be safely neglected in the setup that we considered. Where comparable, our results agree with a previous analysis as well as with experimental measurements at 13 TeV.

\medskip \medskip
This research is supported by grant K 125105 of the National Research, Development and Innovation Office in Hungary.

\end{document}